\documentclass[aps,prb,twocolumn,groupedaddress, amsmath, amssymb]{revtex4-2}

\usepackage{graphicx}  
\usepackage{ulem}
\usepackage{dcolumn}   
\usepackage{bm}        
\usepackage{epstopdf}
\usepackage{subcaption}
\usepackage[colorlinks,linkcolor=blue,citecolor=blue,hyperindex]{hyperref}

\newcommand{\nd}{{\vphantom{\dagger}}}
\newcommand{\HP}{\text{(HP)}}
\newcommand{\Proj}{\text{(P)}}
\newcommand{\GA}{\text{(GA)}}
\newcommand{\Vect}[1]{\boldsymbol{#1}}

\newcommand{\boson}{a}
\newcommand{\pairing}{\Delta}

\newcommand{\LSWTexp}[1]{\langle #1 \rangle_{\text{LSWT}}}
\newcommand{\GAexp}[1]{\langle #1 \rangle_{\text{GA}}}

\newcommand{\CR}[1]{{#1}}

\begin{document}
\title{Projected Holstein-Primakoff boson representation of quantum spins for spin wave theory}

\author{Ke Liu}
\thanks{These authors contribute equally.}
\affiliation{International Center for Quantum Materials, School of Physics, Peking University, Beijing 100871, China}

\author{Fangyu Xiong}
\thanks{These authors contribute equally.}
\affiliation{International Center for Quantum Materials, School of Physics, Peking University, Beijing 100871, China}

\author{Fa Wang}
\email{wangfa@pku.edu.cn}
\affiliation{International Center for Quantum Materials, School of Physics, Peking University, Beijing 100871, China}
\affiliation{Collaborative Innovation Center of Quantum Matter, Beijing 100871, China}

\date{\today}
\begin{abstract}
The Holstein-Primakoff boson representation of quantum spins and associated large-$S$ expansion
have been the standard framework for describing the spin wave excitations in magnetically order phases 
of quantum spin systems. 
However, we will show that the omission of projection operators and normal-ordering 
in this representation can produce incorrect magnon Hamiltonians for finite $S$. 
We will present the exact normal-ordered forms of the finite-$S$ projection operators
and projected Holstein-Primakoff boson representations of spin and quadrupole operators, 
which can produce \textit{exact} two-magnon interaction terms under ferromagnetic or fully polarized states. 
We will also discuss the difficulties of applying this projected representation 
to antiferromagnetic spin wave theory.
\end{abstract}

\maketitle

\section{Introduction}
Many quantum magnets have been proposed as candidates or parent compounds of exotic quantum matter,  
including quantum spin liquids\cite{RevModPhys.89.025003} 
and unconventional superconductors upon doping\cite{RevModPhys.78.17}. 
However, most quantum spin systems in two or higher dimensions are magnetically ordered, 
whose elementary excitations are supposed to be weakly interacting bosonic magnons (quantized spin waves)\cite{ZPhysik.61.206, PhysRev.58.1098, PhysRev.87.568, PhysRev.86.694, PhysRev.102.1217, SovPhysJETP.64.654}.
Theoretical studies of magnons are usually based on the Holstein-Primakoff(HP) boson representation of quantum spins
and the associated large-$S$ expansion of the quantum spin models\cite{PhysRev.58.1098, PhysRev.87.568, PhysRev.117.117}.
The non-interacting parts of these HP boson Hamiltonians form the basis of the linear spin wave theory(LSWT)
and have been widely used to fit the spin excitation spectra of quantum magnets measured by various experiments\cite{RevModPhys.63.1}.

Recently it has been realized that magnon interactions can be important in quantum spin systems.
Theoretically, it will cause decay and dispersion corrections of high energy magnons\cite{RevModPhys.85.219, PhysRevB.71.184440}, 
and may also significantly modify low energy magnon dispersions in frustrated quantum magnets.
For example, it may lift the ``pseudo-Goldstone'' modes
(zero-energy magnon modes in LSWT due to accidental classical ground state degeneracy) 
to finite energy\cite{PhysRevB.46.11137, PhysRevLett.121.237201}.
Magnon interactions may also create magnon bound states\cite{PhysRevLett.11.336, PhysRev.132.85} whose condensation can generate spin nematic orders\cite{PhysRevB.44.4693, PhysRevLett.96.027213, Zhitomirsky_2010, PhysRevB.87.144417}.

On the experimental side, recent developments of spectroscopic probes including 
X-ray and terahertz spectroscopy 
have enabled precise measurements of bimagnon excitations\cite{PhysRevLett.100.097001, PhysRevB.104.144408}.
It is pertinent to have an accurate theory for magnon interaction effects 
in order to understand these experimental developments.

Theories of magnons are mainly 
based on the HP boson representation(indicated by superscript $^\HP$) and associated large-$S$ expansion\cite{PhysRev.117.117},
which replaces spin operators $\hat{S}_a$ by boson operators $\hat{S}^\HP_a$ as follows, 
\begin{subequations}
\begin{align}
\hat{S}_z^\nd\to\hat{S}_{z}^\HP & =S-\hat{n},
\label{equ:HPrep.a}\\
\hat{S}_-^\nd \equiv \hat{S}_x^\nd-i\hat{S}_y^\nd\to \hat{S}_-^\HP & =\hat{\boson}^{\dagger}(2S-\hat{n}^\nd)^{1/2},
\label{equ:HPrep.b}\\
\hat{S}_+^\nd \equiv \hat{S}_x^\nd+i\hat{S}_y^\nd\to \hat{S}_+^\HP & =(2S-\hat{n}^\nd)^{1/2}\hat{\boson}^\nd.
\label{equ:HPrep.c}
\end{align}
\end{subequations}
Here $\hat{\boson}^\nd$ and $\hat{\boson}^\dagger$ are annihilation and creation operators of HP bosons respectively, 
and $\hat{n}^\nd\equiv\hat{\boson}^{\dagger}\hat{\boson}^\nd$ is the HP boson number operator.
The large-$S$ expansion further expands the factor $(2S-\hat{n}^\nd)^{1/2}$ into 
$\sqrt{2S} [1-\sum_{k=1}^{\infty}\frac{(2k-1)!!}{2k!!(2k-1)}(\frac{\hat{n}}{2S})^k ]$, 
so the spin Hamiltonians can be rearranged into different orders of $1/S$.
The spin Hilbert space is mapped to a subspace of bosons as
$|\hat{S}_z=S-n\rangle \to |\hat{n}=n\rangle \equiv \frac{1}{\sqrt{n!}}\hat{\boson}^{\dagger n}|\text{vac}\rangle$
for $0\leq n \leq 2S$, where $|\text{vac}\rangle$ is the HP boson vacuum.

However, \CR{the projection to physical spin Hilbert space is not implemented in typical spin wave theory calculations, 
the truncation of large-$S$ expansion will produce incorrect matrix elements between physical spin states for finite $S$ and the resulting boson Hamiltonian will further mix physical and unphysical states. 
This problem has been noted for a long time\cite{PhysRev.102.1217,RevModPhys.63.1, PhysRev.58.1098, PhysRev.102.1217, SovPhysJETP.64.654}. 
More recently, several works have improved the large-$S$ expansion by infinite\cite{JPhysAMathGen.32.6687} or finite\cite{PhysRevResearch.2.043243, 10.21468/SciPostPhys.10.1.007} normal-ordered series of boson operators, such that the resulting bosonic form of spin operators has correct matrix elements between physical spin states. Some of these works have further studied the effects of these changed spin operator expressions in antiferromagnetic spin wave theory\cite{JPhysAMathGen.32.6687, roberts2025highaccuracyevaluationnonthermalmagnetic} albeit with certain approximations.
In this work,}
we will show that the omission of projection to physical Hilbert space in the original HP boson representation
will produce incorrect magnon interaction terms for finite-$S$ (especially for $S=1/2$),  
which may manifest as spurious multi-magnon states or instabilities. 
This can be fixed by a ``projected HP boson representation'' implementing exact projections.
The resulting infinite series of normal-ordered boson operators 
can in principle produces exact spectrum for physical spin states
and vanishing eigenvalues for unphysical states. 
For ferromagnetic(FM) states, truncation to finite orders of boson operators
can still produce exact spectrum for physical states with a small number of magnons.

This paper is organized as follows.
In Sec.~\ref{sec:SingleSpinOperators} 
we will derive the explicit form the projection and projected spin operators
and discuss the relation and differences between our presentation 
and some previous works\cite{JPhysAMathGen.32.6687, PhysRevResearch.2.043243, 10.21468/SciPostPhys.10.1.007}.
In Sec.~\ref{sec:FM-Applications} and \ref{sec:AFM-Applications} we will
demonstrate the application of our representation
to the FM XXZ model with single-ion anisotropy
and discuss the difficulties of applying this to antiferromagnetic(AFM) states. 
Section~\ref{sec:Conclusion} contains a summary and further discussions.
Technical details are presented in Appendixes.

\section{Projected HP boson representation for single-spin operators}
\label{sec:SingleSpinOperators}
The original HP spin operators Eq.~(\ref{equ:HPrep.a}-\ref{equ:HPrep.c}) 
have correct matrix elements between physical spin-$S$ states. 
However they also have nonzero matrix elements between unphysical HP boson states.
For example, bimagnon states in LSWT contain components with double-occupancy 
$|\hat{n}=2\rangle$ on some sites, which should be projected out for $S=1/2$ systems. 
But the original HP spin operators acting on double-occupancy states produce nonvanishing results even for $S=1/2$.
Therefore the proper representation should be
$\hat{S}_{a}^\Proj = \hat{P}_S^\nd \hat{S}_{a}^\HP \hat{P}_S^\nd$,
where 
$\hat{P}_S^\nd$ is the projection operator for physical spin-$S$ states, 
namely
$\hat{P}_S^\nd|\hat{n}=n\rangle=\left \{\begin{array}{ll}
|\hat{n}=n\rangle, & n\leq 2S;\\
0, & n> 2S.
\end{array}\right .
$
The superscript $^\Proj$ indicates projected operators hereafter.

The projection operator $\hat{P}_S^\nd$ and projected HP spin operators
$\hat{S}_{a}^\Proj$ can be expressed in terms of HP boson operators as infinite series. 
The details of the derivation are given in Appendix~\ref{appsec:Derivation}. 
The results are
\begin{widetext}
\begin{subequations}
\begin{align}
\hat{P}_S^\nd & =\ :e^{-\hat{n}}\sum_{k=0}^{2S}\frac{\hat{n}^k}{k!} :\ 
=1+p_{S,2}\hat{\boson}^{\dagger 2} \hat{\boson}^{\nd 2}+\dots,
\label{equ:ProjRepS.a}\\
\hat{S}_{z}^\Proj & =\ :e^{-\hat{n}}\sum_{k=0}^{2S}\frac{\hat{n}^k (S-k)}{k!} :\ 
= S -\hat{n}+z_{S,2}\hat{\boson}^{\dagger 2} \hat{\boson}^{\nd 2}+\dots,
\label{equ:ProjRepS.b}\\
\hat{S}_+^\Proj & =\ :e^{-\hat{n}}\sum_{k=0}^{2S}\frac{\hat{n}^k\hat{\boson}^\nd\sqrt{2S-k}}{k!} :\ 
=\sqrt{2S}\left [\hat{\boson}^\nd -\left (1-\sqrt{1-\frac{1}{2S}}\right )\hat{\boson}^\dagger \hat{\boson}^{\nd 2}+\dots\right ],
\label{equ:ProjRepS.c}\\
\hat{S}_-^\Proj & =\ :e^{-\hat{n}}\sum_{k=0}^{2S}\frac{\hat{\boson}^\dagger\hat{n}^k\sqrt{2S-k}}{k!} :\ 
=\sqrt{2S}\left [\hat{\boson}^\dagger -\left (1-\sqrt{1-\frac{1}{2S}}\right )\hat{\boson}^{\dagger 2} \hat{\boson}^{\nd}+\dots\right ].
\label{equ:ProjRepS.d}
\end{align}
\end{subequations}
\end{widetext}
Here 
$e^{-\hat{n}}$ should be expanded into $\sum_{k=0}^{\infty} \frac{(-\hat{n})^k}{k!}$, 
and $:\ :$ means normal-ordering of HP boson operators (placing $\hat{\boson}^\dagger$ in front of $\hat{\boson}$). 
For example, $:\hat{n}^k:\ =\hat{\boson}^{\dagger k}\hat{\boson}^{\nd k}
=\prod_{j=0}^{k-1}(\hat{n}-j)$. 
A few terms of these series are also presented, 
where $p_{S,2}=\left \{
\begin{array}{ll}
-1/2, & S=1/2;\\
0, & S>1/2,
\end{array}\right .$
and 
$z_{S,2}=\left \{
\begin{array}{ll}
3/4, & S=1/2;\\
0, & S>1/2,
\end{array}\right .
$
are essential for the hardcore constraint in the spin-$1/2$ case. 
The advantage of this normal-ordered form is that 
$k$th order terms (with $k$ annihilation operators) will not affect states with fewer($<k$) HP bosons.

The projected HP representation Eq.~(\ref{equ:ProjRepS.b}-\ref{equ:ProjRepS.d})
and the original HP representation Eq.~(\ref{equ:HPrep.a}-\ref{equ:HPrep.c}) 
applied to a bilinear spin Hamiltonian
(with only $\hat{S}_{i,a}J_{ij,ab}\hat{S}_{j,b}$ terms) will produce the same LSWT Hamiltonian of non-interacting bosons, 
but the boson interaction terms will be different for any finite $S$.

This projection should also be applied to the HP boson representation 
of any single-spin operator $\hat{f}= f(\hat{\Vect{S}})$.  
Its projected representation $\hat{f}^\Proj
=\hat{P}_S f(\hat{\Vect{S}}^\HP) \hat{P}_S$
can also be expressed as infinite series of normal-ordered terms
(see Appendix~\ref{appsec:Derivation}). 
For example, the projected HP representations of spin quadrupole operators 
for $S\geq 1$ are
\begin{widetext}
\begin{subequations}
\begin{align}
(\hat{S}_z^2)^\Proj & =\ 
:e^{-\hat{n}}\sum_{k=0}^{2S} \frac{\hat{n}^k (S-k)^2}{k!}:\ 
=S^2-(2S-1)\hat{n}^\nd+\hat{\boson}^{\dagger 2} \hat{\boson}^{\nd 2}+\dots
\label{equ:ProjRepQuad.a}
\\
(\hat{S}_-\hat{S}_z+\hat{S}_z\hat{S}_-)^\Proj 
& = \ 
:e^{-\hat{n}}\sum_{k=0}^{2S} \frac{\hat{\boson}^\dagger \hat{n}^k \sqrt{2S-k}(2S-2k-1)}{k!}:\ 
=(2S-1)\sqrt{2S}\hat{\boson}^\dagger+z_{-,S,2}\hat{\boson}^{\dagger 2} \hat{\boson}^{\nd}+\dots
\label{equ:ProjRepQuad.b}
\\
(\hat{S}_-^2)^\Proj & =\ 
:e^{-\hat{n}}\sum_{k=0}^{2S} \frac{\hat{\boson}^\dagger\hat{\boson}^\dagger \hat{n}^k \sqrt{(2S-k)(2S-k-1)}}{k!}:\ 
=\sqrt{2S(2S-1)}\hat{\boson}^{\dagger 2}
+z_{--,S,2}\hat{\boson}^{\dagger 3}\hat{\boson}^\nd
+\dots
\label{equ:ProjRepQuad.c}
\end{align}
\end{subequations}
\end{widetext}
Here $z_{-,S,2}=(2S-3)\sqrt{2S-1}-(2S-1)\sqrt{2S}$
and $z_{--,S,2}=\sqrt{(2S-1)(2S-2)}-\sqrt{2S(2S-1)}$,
and we have omitted $\hat{S}_+\hat{S}_z+\hat{S}_z\hat{S}_+$
and $\hat{S}_+^2$
since they are just the hermitian conjugate of Eq.~(\ref{equ:ProjRepQuad.b},\ref{equ:ProjRepQuad.c}).
For $S=1/2$, these quadrupole operators reduce to
$(\hat{S}_z^2)^\Proj=(1/4)\hat{P}_{1/2}$, 
$(\hat{S}_-\hat{S}_z+\hat{S}_z\hat{S}_-)^\Proj=0$,
and $(\hat{S}_-^2)^\Proj=0$.

To exactly map a spin Hamiltonian $\hat{H}=H(\{\hat{S}_{i,a}\})$ to a boson Hamiltonian, 
we need to perform the full projection $\hat{P}=\prod_{i}\hat{P}_{S,i}$ for all sites $i$, 
on the HP boson representation of the spin Hamiltonian. 
The fully-projected bosonic Hamiltonian $\hat{H}^\Proj
=\hat{P} H(\{ \hat{S}^\HP_{i,a} \}) \hat{P}
=\hat{P} H(\{ \hat{S}^\Proj_{i,a} \}) \hat{P}
$ 
has the exact eigenvalues and eigenvectors for any physical state,
and has vanishing eigenvalue for any unphysical state. 
Unfortunately $\hat{H}^\Proj$ contains infinite-range boson interactions. 
In the same spirit of the treatment for projection in $t$-$J$ models\cite{SupercondSciTechnol.1.36}, 
we \textit{approximate} the fully-projected $\hat{H}^\Proj$ by locally projected terms $\hat{H}^\Proj_{\text{local}}
= \sum_{i} \hat{P}_{S,i} \hat{H}_i^\HP \hat{P}_{S,i}
+\sum_{(i,j)} \hat{P}_{S,i} \hat{P}_{S,j}\hat{H}_{(i,j)}^\HP \hat{P}_{S,i} \hat{P}_{S,j}
+\dots$, 
where $\hat{H}_i^\HP$ is the HP boson representation for a local spin term on site $i$, 
$\hat{H}_{(i,j)}^\HP$ is for a local term on bond $(i,j)$, and so forth.
This local bosonic Hamiltonian $\hat{H}^\Proj_{\text{local}}$
still has the exact eigenvalues and eigenvectors for any physical state,
but may have nontrivial energies for certain unphysical states
(\textit{e.g.} one unphysical site with faraway physical excitations). 
\CR{For example, consider two $S=1/2$ spins $i$ and $j$ with Hamiltonian 
$\hat{H}=-h\hat{S}_{i,z}-h\hat{S}_{j,z}$, 
the fully-projected bosonic Hamiltonian is $\hat{H}^\Proj=-h\hat{S}^\Proj_{i,z}\hat{P}_{1/2,j}
-h\hat{P}_{1/2,j}\hat{S}^\Proj_{j,z}$, while the locally-projected Hamiltonian is $\hat{H}^\Proj_{\text{local}}=-h\hat{S}^\Proj_{i,z}
-h\hat{S}^\Proj_{j,z}$.
The unphysical state $|n_i=2\rangle|n_j=1\rangle$ appears to be a nontrivial excitation of $\hat{H}^\Proj_{\text{local}}$
with excitation energy $h$ (relative to ground state $|n_i=0\rangle|n_j=0\rangle$), although $\hat{S}^\Proj_{i,z}|n_i=2\rangle=0$ vanishes.
}
In the following we will always use this local projection approximation.

We now discuss the relations and differences between our results 
and some previous works in this direction. 
Ref.~\cite{JPhysAMathGen.32.6687} also tried to implement the exact projection
and expressed the spin and quadrupole operators as infinite series of normal-ordered terms, 
but did not get the succinct form and the solution to all series coefficients. 
Ref.~\cite{PhysRevResearch.2.043243} and Ref.~\cite{10.21468/SciPostPhys.10.1.007}
represented $\sqrt{1-\hat{n}/2S}$ and thus $\hat{S}_\pm$ as finite series of normal-ordered terms, 
which had been suggested before\cite{RevModPhys.63.1} albeit without normal-ordering.
These finite series for $\hat{S}_\pm$ match our low order terms 
and have correct matrix elements on physical states with $n\leq 2S$, 
but they also have nontrivial matrix elements on unphysical states. 
\CR{For example, $\hat{S}_-$ for $S=1/2$ is represented by $(\hat{\boson}^\dagger -\hat{\boson}^\dagger\hat{\boson}^\dagger\hat{\boson}^\nd)$
in Ref.~\cite{PhysRevResearch.2.043243, 10.21468/SciPostPhys.10.1.007}, 
this operator acted on unphysical $|n=2\rangle$ state will produce 
$-\sqrt{2}|n=3\rangle$ instead of a vanishing result.
}

All of these works\cite{JPhysAMathGen.32.6687, PhysRevResearch.2.043243, 10.21468/SciPostPhys.10.1.007}
did not provide the exact series for the projection operator $\hat{P}_S$ given here, 
and Ref.~\cite{PhysRevResearch.2.043243, 10.21468/SciPostPhys.10.1.007} 
did not consider the projection effect on $\hat{S}^\HP_z=S-\hat{n}$. 
It will be clear later in Sec.~\ref{sec:FM-Applications} 
that $\hat{P}_S$ and projected $\hat{S}_z^\Proj$
are important for getting the correct boson interaction terms.
For example, the HP representation for an FM Ising coupling term 
$-\hat{S}_{i,z}\hat{S}_{j,z}$ is $-(S-\hat{n}_i)(S-\hat{n}_j)$, whose apparent ground state
would be an unphysical state with boson numbers $n_i,n_j\to +\infty$.  

Another important note to take is 
that the contribution of the single-ion anisotropy term
$\hat{S}_z^2$ to the LSWT Hamiltonian 
is $(1-2S)\hat{n}$ in our representation
and in Ref.~\cite{JPhysAMathGen.32.6687, PhysRevResearch.2.043243}, 
which is different from the large-$S$ expansion result\cite{JPhysCondensMatter.27.166002} 
$(-2S\hat{n})$ due to the ignored normal-ordering of $\hat{n}^2$ term. 
It is important to use the result of $(1-2S)\hat{n}$ in LSWT for finite $S$, 
in order to get correct magnon dispersion and avoid spurious magnon instabilities\cite{JPhysAMathGen.32.6687}.

\section{Applications to ferromagnetic spin wave theory}
\label{sec:FM-Applications}

Consider the nearest-neighbor(NN) XXZ model
with single-ion anisotropy $D_z$ and external field $h$
on square lattice, 
\begin{equation}
\begin{aligned}
\hat{H}_{\text{FM}}=
&
-\sum_{i}[h(\hat{S}_{i,z}-S)+D_z(\hat{S}_{i,z}^2-S)]
\\
&
-
\sum_{<ij>}J_z(\hat{S}_{i,z}\hat{S}_{j,z}-S^2)
\\
&
-\sum_{<ij>}J_\perp (\hat{S}_{i,x}\hat{S}_{j,x}+\hat{S}_{i,y}\hat{S}_{j,y}).
\end{aligned}
\label{equ:H-XXZ-FM}
\end{equation}
The ground state is supposed to be the FM state with $\hat{S}_{i,z}=S$,
corresponding to the HP boson vacuum $|\text{vac}\rangle$. 
The ground state energy has been shifted to zero, so this Hamiltonian only contains supposedly small fluctuations.

As discussed in Sec.~\ref{sec:SingleSpinOperators}, 
we will approximate Eq.~(\ref{equ:H-XXZ-FM}) by the locally projected HP boson Hamiltonian
\begin{equation}
\begin{aligned}
& \hat{H}_{\text{FM}}^\Proj
\\
\approx
&
-\sum_{i}[h(\hat{S}^\Proj_{i,z}-S \hat{P}_{S,i})
+D_z ((\hat{S}_{i,z}^2)^\Proj-S^2\hat{P}_{S,i})]
\\
&
-
\sum_{<ij>}J_z(\hat{S}^\Proj_{i,z}\hat{S}^\Proj_{j,z}-S^2\hat{P}_{S,i}\hat{P}_{S,j})
\\
&
-
\sum_{<ij>}J_\perp
(\hat{S}^\Proj_{i,x}\hat{S}^\Proj_{j,x}+\hat{S}^\Proj_{i,y}\hat{S}^\Proj_{j,y}).
\end{aligned}
\label{equ:H-XXZ-FM-Proj}
\end{equation}

In FM spin wave theory, the HP boson number equals to the magnon number and is conserved. 
For excited states with magnon number $\leq 2S+1$, there is at most one site with $2S+1$ bosons. 
then the local Hamiltonian in Eq.~(\ref{equ:H-XXZ-FM-Proj}) produces the exact energies and wave functions for physical spin states, 
and unphysical states with $2S+1$ bosons on a single site will be eigenstates with exactly vanishing energy independent of model parameters. 
This is guaranteed by shifting the classical energies of every term in the hamiltonian to zero, 
so that projecting out an unphysical site, which effectively creates a vacancy in the lattice, 
does not change the classical ground state energy.
Therefore the local projection operators multiplied by energy shifts
in Eq.~(\ref{equ:H-XXZ-FM-Proj}) are important for correct magnon interactions, 
without them the $(2S+1)$-boson unphysical states would appear as spurious finite energy excitations. 

In contrast, the original HP representation $\hat{H}_{\text{FM}}^{\HP}$  
produces nontrivial eigenvalues for unphysical states,
and the large-$S$ expansion truncated to finite order may
mix physical and unphysical states
and produce incorrect eigenvalues even for physical states. 
The representations in Ref.~\cite{PhysRevResearch.2.043243, 10.21468/SciPostPhys.10.1.007}
can produce the correct eigenvalues for physical states with a few magnons,
but will have nontrivial eigenvalues for unphysical states. 
This is the main advantage of our projected HP representation. 

In the following we will compute bimagnon energies in spin-$1/2$ FM states 
on square lattice by the projected HP representation and the original spin wave theory 
up to $O(S^0)$ order, and compare them to the exact spin Hamiltonian results.

The LSWT Hamiltonian for FM states is
\begin{equation}
\begin{aligned}
\hat{H}_{\text{LSWT,FM}}=
&
\sum_{i}h_{\text{eff}}\hat{n}_i
+\sum_{<ij>}
SJ_z(\hat{n}_i^\nd+\hat{n}_j^\nd)
\\
&
-\sum_{<ij>}
SJ_\perp (\hat{\boson}_i^\dagger\hat{\boson}_j^\nd+\hat{\boson}_j^\dagger\hat{\boson}_i^\nd).
\end{aligned}
\label{equ:LSWT-FM}
\end{equation}
where $h_{\text{eff}}\equiv h+(2S-1)D_z$. 
Note that we have used the proper LSWT treatment for single-ion anisotropy 
discussed in Sec.~\ref{sec:SingleSpinOperators}.

For square lattice, 
$\hat{H}_{\text{LSWT,FM}}=
\sum_{\Vect{k}}
\epsilon_{\text{FM},\Vect{k}} \hat{\boson}_{\Vect{k}}^\dagger\hat{\boson}_{\Vect{k}}^\nd
$,
where the magnon dispersion $\epsilon_{\text{FM},\Vect{k}}$ is
\begin{equation}
\epsilon_{\text{FM},\Vect{k}}
=h_{\text{eff}}+S[4J_z-2J_\perp (\cos k_x + \cos k_y)],
\label{equ:epsilon_FM}
\end{equation}
Here 
$\hat{\boson}_{\Vect{k}}^\nd=N_{\text{site}}^{-1/2}\sum_{\Vect{r}} \hat{\boson}_{\Vect{r}}^\nd e^{-i\Vect{k}\cdot \Vect{r}}$
is the annihilation operator for magnon with momentum $\Vect{k}$.
This magnon dispersion is exactly the eigenvalue of plane waves of single spin flips
$\sum_{\Vect{r}}e^{i\Vect{k}\cdot\Vect{r}}\hat{S}_{\Vect{r},-}|\text{vac}\rangle$. 
Stability of the FM state requires that $\epsilon_{\text{FM},\Vect{k}}\geq 0$ for any $\Vect{k}$, 
namely $h_{\text{eff}}+4SJ_z-4S|J_\perp|\geq 0$.

In LSWT, bimagnons of momentum $\Vect{q}$ have continuous energies 
$\epsilon_{\text{LSWT,FM},\Vect{q},\Vect{k}}=\epsilon_{\text{FM},\Vect{k}}+\epsilon_{\text{FM},\Vect{q}-\Vect{k}}$
bounded by 
$2h_{\text{eff}}+8SJ_z\mp 4S|J_\perp|(|\cos(q_x/2)|+|\cos{q_y/2}|)$. 
Magnon interactions will modify this spectrum and may create bimagnon bound states out of this continuum.

The original HP representation expanded to $O(S^0)$ order contains the following magnon interaction terms,
\begin{equation}
\begin{aligned}
\hat{H}^{\HP,(2)}_{\text{FM}} = &
\frac{J_\perp}{4} \sum_{<\Vect{r}\Vect{r}'>}
(
\hat{\boson}_{\Vect{r}'}^{\dagger 2} \hat{\boson}_{\Vect{r}'}^{\nd}\hat{\boson}_{\Vect{r}}^{\nd}
+
\hat{\boson}_{\Vect{r}'}^\dagger \hat{\boson}_{\Vect{r}}^\dagger \hat{\boson}_{\Vect{r}}^{\nd 2}
+\text{h.c.})
\\
& 
-D_z \sum_{\Vect{r}} \hat{n}_{\Vect{r}}^2
-J_z \sum_{<\Vect{r}\Vect{r}'>}
\hat{n}_{\Vect{r}}^\nd \hat{n}_{\Vect{r}'}.
\end{aligned}
\label{equ:FM-Interactions-HP}
\end{equation}
Here $\text{h.c.}$ is hermitian conjugate of the previous terms. 
Note that the $D_z$ term here should actually be normal-ordered,
and is \textit{incorrectly} present even for spin-$1/2$ case.

The projected HP representation in Eq.~(\ref{equ:ProjRepS.a}-\ref{equ:ProjRepS.d}, \ref{equ:ProjRepQuad.a}) produces different two-magnon interaction terms,
\begin{equation}
\begin{aligned}
& \hat{H}^{\Proj,(2)}_{\text{FM}}
\\
= & 
J_\perp S\sum_{<\Vect{r}\Vect{r}'>}
\left ( 1-\sqrt{1-\frac{1}{2S}} \right )
(
\hat{\boson}_{\Vect{r}'}^{\dagger 2} \hat{\boson}_{\Vect{r}'}^{\nd}\hat{\boson}_{\Vect{r}}^{\nd}
+
\hat{\boson}_{\Vect{r}'}^\dagger \hat{\boson}_{\Vect{r}}^\dagger \hat{\boson}_{\Vect{r}}^{\nd 2}
+\text{h.c.})
\\
&
-\sum_{\Vect{r}}
[(4SJ_z+h) (z_{S,2}-S p_{S,2})+(1-\delta_{S,1/2})D_z] \hat{\boson}_{\Vect{r}}^{\dagger 2} \hat{\boson}_{\Vect{r}}^{\nd 2}
\\
&
-
J_z \sum_{<\Vect{r}\Vect{r}'>}
\hat{n}_{\Vect{r}}^\nd \hat{n}_{\Vect{r}'}.
\end{aligned}
\label{equ:FM-Interactions-P}
\end{equation}
Here $z_{S,2}-S p_{S,2}=\delta_{S,1/2}=1$ for spin-$1/2$
and vanishes for higher $S$. 
As discussed above, $\hat{H}_{\text{LSWT,FM}}+\hat{H}^{\Proj,(2)}_{\text{FM}}$ 
has the exact bimagnon energies and wave functions in addition to zero-energy unphysical double-occupancy states, 
because the omitted normal-ordered terms in $\hat{H}_{\text{FM}}^\Proj$ vanish on any states with two or less HP bosons.

To illustrate the difference between these results, 
the bimagnon energies for $S=1/2$ case on a $12\times 12$ square lattice with periodic boundary  
are computed for 
the projected HP representation $\hat{H}_{\text{LSWT,FM}} +\hat{H}^{\Proj,(2)}_{\text{FM}}$, 
the HP represention $\hat{H}_{\text{LSWT,FM}} +\hat{H}^{\HP,(2)}_{\text{FM}}$,
the HP representation with $D_z$ term removed, 
and the exact spin Hamiltonian, 
respectively. 
The details of the derivation are given in 
Appendix~\ref{appsec:BimagnonSpinHamiltonian} and \ref{appsec:BimagnonBosonHamiltonian}. 
The results for some $\Vect{q}$ are shown in Fig.~\ref{fig:FM-bimagnon}, 
and source codes of computation are available in Supplemental Material\footnote{See Supplemental Material at [URL] for the source codes of computation shown in Fig.~\ref{fig:FM-bimagnon}.}.
For any $\Vect{q}$, the projected HP representation
produces the exact bimagnon energies(which are independent of $D_z$ as expected) 
in addition to a single unphysical zero-energy state
corresponding to plane waves of double-occupancy, 
$\sum_{\Vect{r}}e^{i\Vect{k}\cdot \Vect{r}}\hat{\boson}_{\Vect{r}}^{\dagger 2}|\text{vac}\rangle$.
The original HP representation results 
contain a spurious bimagnon bound state for large $D_z$, 
due to the $-D_z\hat{n}_i^2$ term which should be absent for spin-$1/2$ cases.
If we remove this $D_z$ term by hand from the original HP representation, 
the unphysical state will merge into the continuum, 
and slightly change the physical spectrum. 

\begin{figure*}
\includegraphics[scale=0.65]{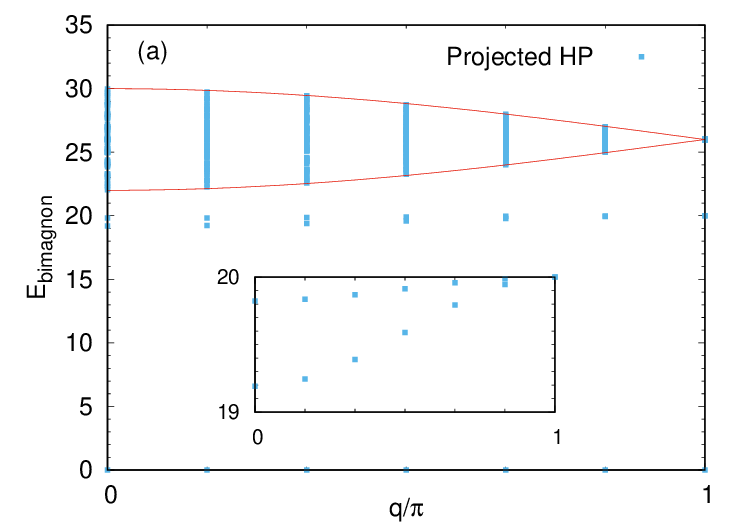}
\includegraphics[scale=0.65]{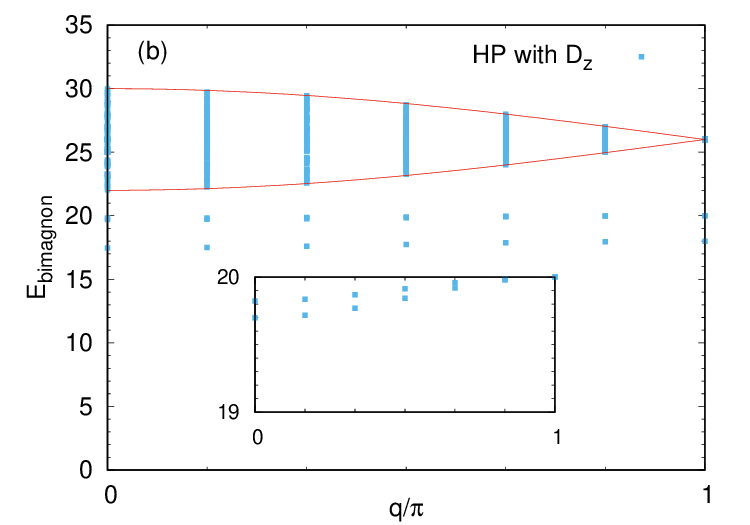}
\\
\includegraphics[scale=0.65]{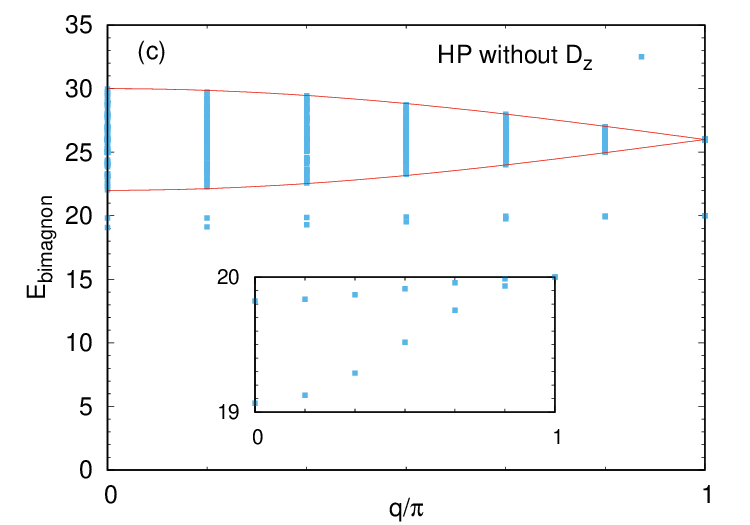}
\includegraphics[scale=0.65]{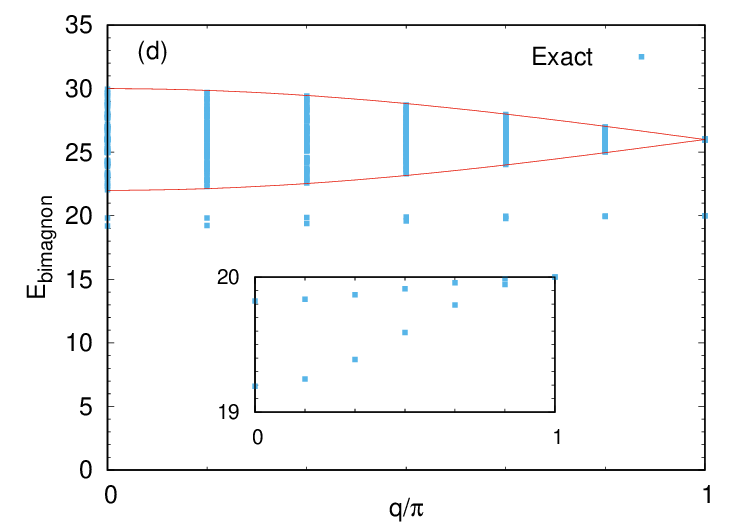}
\caption{
Calculated bimagnon energies at momentum $\Vect{q}=(q,q)$ for spin-$1/2$ FM XXZ model 
Eq.~(\ref{equ:H-XXZ-FM}) on $12\times 12$ square lattice  
with parameters $h=1$, $J_z=6$, $J_\perp=1$, and $D_z=4$. 
Lines in figures are upper and lower bounds of the bimagnon continuum in LSWT.
Insets are detailed dispersions of the two bimagnon bound states below the continuum. 
See Appendix~\ref{appsec:BimagnonSpinHamiltonian} and \ref{appsec:BimagnonBosonHamiltonian} for calculation details. 
(a) Results using projected HP representation Eq.~(\ref{equ:FM-Interactions-P}).
Note the zero-energy unphysical states. 
(b) Results using original HP representation Eq.~(\ref{equ:FM-Interactions-HP}). 
The lowest energy bimagnon state is spurious
(with more than $70$\% spectral weight in unphysical double-occupancy state). 
(c) Results using original HP representation with $D_z$ term removed. 
The unphysical state merges into the continuum and slightly changes the spectrum. 
(d) Exact bimagnon energies from the spin model Eq.~(\ref{equ:H-XXZ-FM}).
}
\label{fig:FM-bimagnon}
\end{figure*}

\section{Potential applications to antiferromagnetic spin wave theory}
\label{sec:AFM-Applications}
Spin wave theory in AFM states is much more complex than FM cases. 
In Subsec.~\ref{subsec:Recall} we will recall some results and problems of 
AFM (non-)linear spin wave theory, 
and then show some attempted applications of the projected HP 
representation to AFM cases
in Subsec.~\ref{subsec:Attempts}.

\subsection{Difficulties with antiferromagnetic spin wave theory}
\label{subsec:Recall}

Consider the N\'eel order on a bipartite Archimedean lattice, 
with classical ground state $\hat{S}_{i,z}=\pm S$ for 
the $A$,$B$ sublattices respectively.
To apply the HP boson represenation, 
the spin axes on $B$ sublattice have to be rotated 
so that all classical spins are polarized along $+z$ axis\cite{PhysRev.87.568}.
Choose the rotations on $B$ sublattice sites $j$ as
$\hat{S}_{j,z}\to -\hat{S}_{j,z}$, 
$\hat{S}_{j,x}\to -\hat{S}_{j,x}$,
and $\hat{S}_{j,y}\to \hat{S}_{j,y}$, 
then the spin-$1/2$ NN AFM XXZ model without external field is 
\begin{equation}
\hat{H}_{\text{AFM}}
=
-\sum_{<ij>}[J_z (\hat{S}_{i,z}\hat{S}_{j,z}-S^2)
+J_\perp (\hat{S}_{i,x}\hat{S}_{j,x}-\hat{S}_{i,y}\hat{S}_{j,y})].
\label{equ:H-AFM}
\end{equation}
Similar to the FM case, 
the locally projected HP boson Hamiltonian for Eq.~(\ref{equ:H-AFM}) is
\begin{equation}
\begin{aligned}
\hat{H}_{\text{AFM}}^\Proj
\approx
&
-\sum_{<ij>}[J_z (\hat{S}_{i,z}^\Proj\hat{S}_{j,z}^\Proj-S^2\hat{P}_{S,i}\hat{P}_{S,j})
\\
&
+J_\perp (\hat{S}_{i,x}^\Proj\hat{S}_{j,x}^\Proj-\hat{S}_{i,y}^\Proj\hat{S}_{j,y}^\Proj)].
\end{aligned}
\label{equ:H-AFM-P}
\end{equation}

The LSWT Hamiltonian for this N\'eel state is 
\begin{equation}
\hat{H}_{\text{LSWT,AFM}}
=S\sum_{<ij>}[
J_z(\hat{n}_i^\nd+\hat{n}_j^\nd)
-J_\perp (\hat{\boson}_i^\dagger\hat{\boson}_j^\dagger+\hat{\boson}_j^\nd\hat{\boson}_i^\nd)].
\label{equ:LSWT-AFM}
\end{equation}
Its ground state energy per bond is\cite{PhysRevB.43.8321, PhysRevB.44.11869}, 
\begin{equation*}
E_{\text{GS,LSWT,bond}}=2S(J_z \LSWTexp{ \hat{n} }-J_\perp \LSWTexp{ \hat{\boson}_i^\nd\hat{\boson}_j^\nd }).
\end{equation*}
Here $\langle\cdots \rangle_{\text{LSWT}}$ is the expectation value under LSWT ground state $|\Psi_{\text{LSWT}}\rangle$ of Eq.~(\ref{equ:LSWT-AFM}), 
the HP boson density $\LSWTexp{ \hat{n} }$ should be uniform on all sites, 
and the NN pairing amplitude $\LSWTexp{ \hat{\boson}_i^\nd\hat{\boson}_j^\nd }$ should be real and uniform on all NN bonds $<ij>$
for bipartite Archimedean lattices.
For notation simplicity, we define $\bar{n}$ and $\bar{\pairing}$ as
\[
\bar{n}\equiv \LSWTexp{ \hat{n} },\quad \bar{\pairing}=\LSWTexp{ \hat{\boson}_i^\nd\hat{\boson}_j^\nd }. 
\] 
Since the ground state energy of Eq.~(\ref{equ:LSWT-AFM}) should be negative, 
LSWT for Heisenberg models have $\bar{\pairing} > \bar{n}$ 
(see caption of Table~\ref{tab:Comparison}).

The $O(S^0)$ terms in the large-$S$ expansion of HP representation
consist of the following interaction terms
\begin{equation}
\hat{H}_{\text{AFM}}^{\HP,(2)}
=\sum_{<ij>}[-J_z\hat{n}_i^\nd\hat{n}_j^\nd
+\frac{J_\perp}{4}
(
\hat{\boson}_{j}^{\dagger} \hat{\boson}_{j}^{\nd 2}\hat{\boson}_{i}^{\nd}
+
\hat{\boson}_{j}^\dagger \hat{\boson}_{i}^{\dagger 2} \hat{\boson}_{i}^{\nd}
+\text{h.c.})].
\label{equ:AFM-Interactions-HP}
\end{equation}
The 2nd-order non-linear spin wave theory(NLSWT) considers 
the one-loop correction from $\hat{H}_{\text{AFM}}^{\HP,(2)}$ \cite{PhysRevB.43.8321, PhysRevB.44.11869, PhysRevB.46.11137}.
Since the bare two-boson interaction is instantaneous, 
this one-loop approximation is effectively a mean-field treatment of the interaction terms,
which can be more efficiently done in the real-space form.
For N\'eel order on bipartite Archimedean lattice, 
this just changes the LSWT coefficients in Eq.~(\ref{equ:LSWT-AFM}) to
\begin{subequations}
\begin{eqnarray}
SJ_z^{\text{(NLSW)}} & = & SJ_z +J_\perp \bar{\pairing} - J_z \bar{n},
\label{equ:NLSWT-Jz}
\\
SJ_\perp^{\text{(NLSW)}} & = & SJ_\perp  + J_z \bar{\pairing} - J_\perp \bar{n}.
\label{equ:NLSWT-Jp}
\end{eqnarray}
\end{subequations}
For Heisenberg models, this correction increases spin wave velocity\cite{PhysRevB.71.184440}
by a factor of $1+(1/S)(\bar{\pairing} - \bar{n})$, 
while keeping the gapless Goldstone modes (keeping $J_z=J_\perp$). 
The 2nd-order NLSWT correction to the ground state energy 
is basically the expectation value of Eq.~(\ref{equ:AFM-Interactions-HP})
under the (corrected) LSWT, the result per bond is\cite{PhysRevB.43.8321, PhysRevB.44.11869}
\begin{equation}
E_{\text{GS,NLSWT,bond}}^{(2)}=
-J_z(\bar{n}^2+\bar{\pairing}^2)+2J_\perp \bar{n}\bar{\pairing}.
\label{equ:E-GS-NLSWT}
\end{equation}

In easy-axis XXZ model($J_z>J_\perp$), LSWT magnon gap is\cite{PhysRevB.43.8321, PhysRevB.44.11869} 
$zS \sqrt{J_z^2-J_\perp^2}$, where $z$ is the coordination number.
The 2nd-order NLSWT correction Eq.~(\ref{equ:NLSWT-Jz},\ref{equ:NLSWT-Jp}) reduces this gap to
\footnote{The 2nd-order NLSWT magnon gap in
Ref.~\cite{PhysRevB.43.8321, PhysRevB.44.11869} 
are effectively $z(S-\bar{\pairing}+\bar{n})\sqrt{J_z^2-J_\perp^2}$
and different from our result here, 
the reason is likely due to their omission 
of magnon's anomalous self-energy, namely pairing terms, 
generated by interactions, which will further modify magnon dispersions.
} 
$z
\sqrt{(J_z^2-J_\perp^2)
[(S - \bar{n})^2-\bar{\pairing}^2]
}$.

The linear and non-linear spin wave theories have been particularly successful 
in describing AFM orders of two-dimensional quantum spin models\cite{RevModPhys.63.1, RevModPhys.85.219}. 
But they may also produce spurious results when quantum fluctuation is very strong 
due to small coordination number and/or geometric frustration. 
For example, the LSWT for AFM Shastry-Sutherland model\cite{PhysicaBC.108.1069}
incorrectly predicted instability of magnetic order when second-nearest-neighbor coupling equals to 
the NN coupling\cite{EurophysLett.35.145}, which can be fixed by self-consistent NLSWT\cite{PhysRevB.109.134409}.
Another example is the NLSWT for honeycomb XXZ model near the Heisenberg limit,
in this case $\bar{n} + \bar{\pairing} \sim 0.62 > 1/2$ (see Table~\ref{tab:Comparison}), 
so the 2nd-order NLSWT produces spurious magnon gap closing for spin-$1/2$ honeycomb easy-axis XXZ model near the Heisenberg limit
(see Appendix.~\ref{appsec:NLSWTHoneycomb}), which cannot be fixed even by self-consistent NLSWT. 

It is tempting to apply the projected HP representation
to the AFM spin wave theory, 
which unfortunately will encounter the following difficulties. 

Firstly, 
the $J_\perp$ terms in Eq.~(\ref{equ:H-AFM}) break HP boson number conservation,
the ground state is no longer the HP boson vacuum,
the magnon (Bogoliubov quasi-particle of HP boson) number does not equal to the HP boson number,
and magnon interaction terms will further break the magnon number conservation\cite{RevModPhys.85.219}.
Therefore, the projected HP representations are not normal-ordered in terms of magnon operators, 
and truncation to finite order will not produce exact Hamiltonians for a few magnons in AFM states. 
This is generally true when the magnetic order parameter is not a conserved quantity.

Secondly, 
perturbative treatment of boson interactions from the projected HP representation
usually breaks the global $SU(2)$ symmetry of the Heisenberg models
and spoils the gaplessness of Goldstone modes. 
For example, the projected HP representation Eq.~(\ref{equ:H-AFM-P}) truncated to 2nd-order 
produces the following interactions 
\begin{equation}
\begin{aligned}
&
\hat{H}_{\text{AFM}}^{\Proj,(2)}
\\
=
&
-\sum_{<ij>}J_z
[(Sz_{S,2}-S^2p_{S,2})
(\hat{\boson}_{i}^{\dagger 2} \hat{\boson}_{i}^{\nd 2}
+\hat{\boson}_{j}^{\dagger 2} \hat{\boson}_{j}^{\nd 2})
+\hat{n}_i^\nd\hat{n}_j^\nd]
\\
&
+
J_\perp S\sum_{<ij>}
(1-\sqrt{1-\frac{1}{2S}})
(
\hat{\boson}_{j}^{\dagger} \hat{\boson}_{j}^{\nd 2}\hat{\boson}_{i}^{\nd}
+
\hat{\boson}_{j}^\dagger \hat{\boson}_{i}^{\dagger 2} \hat{\boson}_{i}^{\nd}
+\text{h.c.}),
\end{aligned}
\label{equ:AFM-Interactions-P}
\end{equation}
and a mean-field treatment of this 
will in general not preserve $J_z=J_\perp$ for Heisenberg models.

In this sense, the projected HP representation has no apparent advantage over the original representation
in AFM non-linear spin wave theory, because both representations have to be treated with uncontrolled approximations for finite-$S$. 

\subsection{Gutzwiller approximation for projected HP representation in spin-1/2 N\'eel states}
\label{subsec:Attempts}

Despite the difficulties of applying the projected representation 
(and similar representations beyond large-$S$ approximation) 
to the AFM spin wave theory,  
we will show some preliminary results for mean-field and Gutzwiller approximation of the projection effects
for N\'eel states of spin-$1/2$ NN AFM Heisenberg models on bipartite Archimedean lattices.

Firstly, the projected HP representation changes the form of observables 
and thus changes their expectation values even under LSWT ground states. 
For N\'eel states of spin-$1/2$ NN AFM Heisenberg models on bipartite Archimedean lattices, 
the 2nd-order projected HP representation result for staggered moment is 
\begin{equation}
m_s^{\Proj,(2)}= 1/2-\bar{n}+(3/2)\bar{n}^2,
\label{equ:ms-Proj2nd}
\end{equation}
which is larger than the LSWT result\cite{PhysRevB.43.8321}, 
$m_{s,\text{LSWT}}=1/2-\bar{n}$. 
The 2nd-order NLSWT results of $m_s$ for these models are the same as the LSWT results\cite{PhysRevB.43.8321}. 

The 2nd-order interactions from the projected HP representation Eq.~(\ref{equ:AFM-Interactions-P}) 
generate correction to the ground state energy per bond as
\begin{equation}
E^{\Proj,(2)}_{\text{GS,bond}}
=
-J_z (3\bar{n}^2+\bar{\pairing}^2)
+4J_\perp \bar{n}\bar{\pairing},
\label{equ:E-GS-Proj2nd}
\end{equation}
which is different from the NLSWT result Eq.~(\ref{equ:E-GS-NLSWT}).

Similar to the Gutzwiller approximation for $t$-$J$ models\cite{SupercondSciTechnol.1.36},
we also consider a Gutzwiller approximation(GA) for projected LSWT ground states $\hat{P}|\Psi_{\text{LSWT}}\rangle$. 
By omitting correlations between projection operators on different sites, 
the expectation value of a local operator $\hat{O}$ under projected LSWT ground state, 
$\frac{\langle \psi_{\text{LSWT}}|\hat{P}\hat{O}\hat{P}|\psi_{\text{LSWT}}\rangle}{\langle \psi_{\text{LSWT}}|\hat{P}|\psi_{\text{LSWT}}\rangle}$, 
is approximated by its ``GA expectation value'' $\GAexp{\hat{O}}$, 
which replaces the full projection operator $\hat{P}$ by local projection operator $\hat{P}_{\text{local}}$ on those sites involved in $\hat{O}$, 
\begin{equation}
\GAexp{\hat{O}} \equiv \frac{\langle \psi_{\text{LSWT}}|\hat{O}^\Proj|\psi_{\text{LSWT}}\rangle}{\langle \psi_{\text{LSWT}}|\hat{P}_{\text{local}}|\psi_{\text{LSWT}}\rangle}.
\label{equ:P-GA}
\end{equation}
Here the projected operator $\hat{O}^\Proj$ is $\hat{O}^\Proj\equiv \hat{P}_{\text{local}}\hat{O}\hat{P}_{\text{local}}$.

The expectation value of any bosonic operator under the LSWT ground state
can be computed by the Wick expansion 
in terms of $\LSWTexp{\hat{\boson}_i^\dagger \hat{\boson}_j^\nd}$ and
$\LSWTexp{\hat{\boson}_i^\nd \hat{\boson}_j^\nd}$.
Because of the staggered $U(1)$ symmetry of Eq.~(\ref{equ:LSWT-AFM}), 
$\LSWTexp{\hat{\boson}_i^\nd \hat{\boson}_j^\nd}
=0$ if $i,j$ are in the same sublattice, 
and
$\LSWTexp{\hat{\boson}_i^\dagger \hat{\boson}_j^\nd}
=0$ if $i,j$ are in different sublattices.
In the following only two of these correlators, 
the ``HP boson density'' $\bar{n}\equiv \LSWTexp{\hat{\boson}_i^\dagger \hat{\boson}_i^\nd}$
and the ``NN pairing amplitude'' $\bar{\pairing}=\LSWTexp{\hat{\boson}_i^\nd\hat{\boson}_j^\nd}$
for NN bond $<ij>$ 
are needed. 
The calculation details for the various LSWT expectation values are given in Appendix~\ref{appsec:LSWTExpectationValue}.  

The staggered moment $m_s$ 
under Gutzwiller approximation is 
\begin{equation}
m_s^{\GA}
=\frac{\LSWTexp{ \hat{S}_{i,z}^\Proj } }{\LSWTexp{ \hat{P}_{1/2,i} } }
=\frac{1}{2(1+\bar{n})}.
\label{equ:ms-GA}
\end{equation}
which is larger than the LSWT result $m_{s,\text{LSWT}}=1/2-\bar{n}$. 
This is consistent with the intuition that the projection 
suppresses boson number fluctuation and thus reduces the reduction of ordered moments. 

The Gutzwiller approximation of ground state energy of Eq.~(\ref{equ:H-AFM})
involves local projection operator $\hat{P}_{1/2,i}\hat{P}_{1/2,j}$ on bond $<ij>$. 
The final result for the GA of ground state energy per bond of Eq.~(\ref{equ:H-AFM}) is,
\begin{equation}
\begin{aligned}
& E_{\text{GS,bond}}^\GA
\\
=
&
\frac{
[J_z (\bar{n}+\bar{n}^2-\bar{\pairing}^2)
-J_\perp \bar{\pairing}]
[(1+\bar{n})^2+\bar{\pairing}^2]
}
{
(1+\bar{n})^2+\bar{\pairing}^2
+
4[\bar{n}(1+\bar{n})-\bar{\pairing}^2][(1+\bar{n})^2-\bar{\pairing}^2]
}.
\end{aligned}
\label{equ:E-GA}
\end{equation}

\begin{table*}
\begin{tabular}{|l|l|l|l|l|l|}
\hline\hline
 & LSWT  & NLSWT & 2nd-order Proj. & Gutzwiller Approx. & Numerics \\
\hline\hline
Square $m_s$ & $0.3034$ \cite{RevModPhys.63.1} & $0.3034$ \cite{RevModPhys.63.1} & $0.361$ & $0.358$ & $0.307$ \cite{PhysRevB.56.11678} 
\\
Square $E_{\text{GS}}$ & $-0.658$ \cite{RevModPhys.63.1} & $-0.670$ \cite{PhysRevB.43.8321} &  $-0.608$ & $-0.633$ & $-0.669$ \cite{PhysRevB.56.11678}
\\
\hline\hline
Honeycomb $m_s$ & $0.2418$ \cite{PhysRevB.44.11869} & $0.2418$ \cite{PhysRevB.44.11869} & $0.341$ & $0.329$ & $0.27$ \cite{PhysRevB.84.094424, PhysRevLett.110.127205, PhysRevB.88.165138}
\\
Honeycomb $E_{\text{GS}}$ & $-0.532$ \cite{PhysRevB.44.11869} & $-0.549$ \cite{PhysRevB.44.11869} & $-0.468$ & $-0.506$ & $-0.54$ \cite{PhysRevB.84.094424}
\\
\hline
\hline
\end{tabular}
\caption{
Comparison between the values of staggered moment $m_s$ and ground state energy per site $E_{\text{GS}}$ 
in LSWT, 
2nd-order NLSWT, 
2nd-order projected representation 
and Gutzwiller approximation under LSWT ground states, 
and some numerical computations,
for spin-$1/2$ NN Heisenberg AFM models ($J_z=J_\perp=1$) on square and honeycomb lattices. 
Note that the shift of the classical ground state energy in Eq.~(\ref{equ:H-AFM}) has been restored. 
The projected representation and Gutzwiller approximation results are derived from 
Eq.~(\ref{equ:E-GS-NLSWT},\ref{equ:ms-Proj2nd},\ref{equ:E-GS-Proj2nd},\ref{equ:ms-GA},\ref{equ:E-GA}), 
and the LSWT expectation values for HP boson density $\bar{n}\equiv\LSWTexp{ \hat{n} }$
and NN boson pairing amplitude $\bar{\pairing}\equiv \LSWTexp{ \hat{\boson}_i^\nd\hat{\boson}_j^\nd }$. 
For square lattice, 
$\bar{n}\approx 0.1966$ and 
$\bar{\pairing}\approx 0.2756$ \cite{PhysRev.87.568}.
For honeycomb lattice, 
$\bar{n}\approx 0.2582$ and 
$\bar{\pairing}\approx 0.3631$ \cite{PhysRevB.44.11869}.
}
\label{tab:Comparison}
\end{table*}

Table~\ref{tab:Comparison} shows the comparison between LSWT, 2nd-order NLSWT, 2nd-order projected representation
and Gutzwiller approximation under LSWT ground state, and some numerical results
for the staggered moment and ground state energy of square and honeycomb spin-$1/2$ Heisenberg models.
It seems that the projected HP representation under these two approximation schemes 
does not provide a better approximation compared to the (non-)linear spin wave theory
for these two lattices. It might be interesting to explore other models 
with stronger quantum fluctuations, in order to see whether the projected representation performs better there.

\CR{A previous work\cite{roberts2025highaccuracyevaluationnonthermalmagnetic} 
applied the finite series form of $\hat{S}_{\pm}$ operators
in Ref.~\cite{PhysRevResearch.2.043243, 10.21468/SciPostPhys.10.1.007} to study a spin-1/2 chain with XXZ and Dzyaloshinskii-Moriya interactions.
Because the omission of projection effects on $\hat{S}_z$ and the projection operators for ground state energy shift, 
the boson interaction terms from XXZ terms obtained in Ref.~\cite{roberts2025highaccuracyevaluationnonthermalmagnetic}
is slightly different from the boson interaction terms obtained here in Eq.~(\ref{equ:AFM-Interactions-P}).
}


\section{Discussion and Conclusion}
\label{sec:Conclusion}
In summary, 
we have developed a ``projected Holstein-Primakoff boson representation'' of quantum spins
which takes the projection to physical spin space into account exactly. 
We demonstrated its superiority over the original HP representation in 
the calculation of bimagnon energies in ferromagnetic states, 
discussed the difficulties of applying it to antiferromagnetic states 
and showed some preliminary Gutzwiller approximation results. 
For FM states (or states polarized by external field), 
our projected representation truncated to 2nd order can produce the exact bimagnon Hamiltonian. 
Our current methods for applying the projected representation in AFM spin wave theory somehow did not produce more accurate 
results than the linear spin wave theory of square and honeycomb Heisenberg models, 
in comparison to numerical results. 
Nonetheless our technical results here may be helpful for future theoretical studies 
and hopefully will result in more accurate approximations 
for other models with larger quantum fluctuations.

Some early works\cite{PhysRev.102.1217,RevModPhys.63.1} also 
tried to address the problem of omitted projection in bosonic representations of spins,
including the HP representation\cite{PhysRev.58.1098} and the Dyson-Maleev representation\cite{PhysRev.102.1217, SovPhysJETP.64.654}.
However, they all involve non-orthogonal basis and therefore additional approximations. 
There are recent works that tried to represent the spin operators as infinite\cite{JPhysAMathGen.32.6687} or finite series\cite{PhysRevResearch.2.043243, 10.21468/SciPostPhys.10.1.007} of normal-ordered boson operators. 
Ref.~\cite{JPhysAMathGen.32.6687} implemented the exact projection,  
but did not provide general solutions to the series coefficients, 
and only considered application to linear spin wave theory. 
Ref.~\cite{PhysRevResearch.2.043243, 10.21468/SciPostPhys.10.1.007, roberts2025highaccuracyevaluationnonthermalmagnetic} 
obtained the same low order terms for $\hat{S}_\pm$ as the results here and in Ref.~\cite{JPhysAMathGen.32.6687}
\CR{and tried to apply these changed form of spin operators to AFM spin wave theory,}
but they did not consider the projection effect on $\hat{S}_z$. 
None of these works\cite{JPhysAMathGen.32.6687, PhysRevResearch.2.043243, 10.21468/SciPostPhys.10.1.007} derived the explicit form of the projection operator or realize the importance of the projection operator 
in obtaining accurate boson interaction terms.

One important remaining technical problem  
is to preserve the global $SU(2)$ symmetry for AFM Heisenberg models 
under this projected representation. 
This symmetry, namely $[\sum_{i} \hat{S}_{i,y}, \hat{H}_{\text{AFM}}]=0$, 
is preserved order by order in the large-$S$ expansion, 
which maintains the gaplessness of the Goldstone modes. 
It might be fruitful to combine the large-$S$ expansion 
and the projected representation to create a better approximation scheme.

\section{ACKNOWLEDGEMENTS}
FW acknowledges support from National Natural Science Foundation of China (Grant No. 12274004 and No. 11888101).


\appendix

\section{Derivation of the projected HP boson representation for single-spin operators}
\label{appsec:Derivation}

In this Section we derive the explicit infinite series 
for the projected HP representations of single-spin operators shown in main text.

A diagonal operator conserving $\hat{S}_z$ 
is generally of the form $f(\hat{S}_z)$
and should also conserve HP boson number. 
So its projected representation $\hat{f}^\Proj=\hat{P}_S f(S-\hat{n}) \hat{P}_S$
can only contain 
$:\hat{n}^k:\,\equiv \hat{\boson}^{\dagger k} \hat{\boson}^k=
\prod_{j=0}^{k-1}(\hat{n}-j)$,
and $\hat{\boson}^{\dagger k} \hat{\boson}^k|\hat{n}=n\rangle
=\frac{n!}{(n-k)!}|\hat{n}=n\rangle$. 
Assume that $\hat{f}^\Proj=\sum_{k=0}^{\infty}F_{k}\hat{\boson}^{\dagger k} \hat{\boson}^k$, 
then the condition 
\[
\hat{f}^\Proj|\hat{n}=n\rangle=\left \{\begin{array}{ll}
f(S-n)|\hat{n}=n\rangle, & n\leq 2S;\\
0, & n> 2S,
\end{array}\right .
\]
becomes
\[
\sum_{k=0}^{n} \frac{n!}{(n-k)!}F_{k}
=\left \{\begin{array}{ll}
f(S-n), & n\leq 2S;\\
0, & n> 2S.
\end{array}\right .
\]
To solve $F_{k}$, define the generating function $F(t)=\sum_{k=0}^{\infty} F_{k} t^k$, 
the above linear equation of $F_{k}$ 
is the $t^n$ coefficient of the equation 
$e^{t} F(t)=\sum_{k=0}^{2S}\frac{t^k}{k!}f(S-k)$, 
then $F(t)=e^{-t}\sum_{k=0}^{2S}\frac{t^k}{k!}f(S-k)$, 
and 
\[
\hat{f}^\Proj=\ 
:F(\hat{n}):\ 
=\ 
:e^{-\hat{n}}\sum_{k=0}^{2S}\frac{\hat{n}^k}{k!}f(S-k):\ .
\]
The results of $\hat{P}_S$ and $\hat{S}_z^\Proj$ and $(\hat{S}_z^2)^\Proj$ correspond to 
$f(S-n)=1$ and $(S-n)$ and $(S-n)^2$, respectively.

Operators that change $S_z$ quantum number by $-1$
are generally of the form $\hat{f}_-=\hat{S}_- f(\hat{S}_z)$. 
Then the projected $(\hat{f}_-)^\Proj$ should be 
$(\hat{f}_-)^\Proj=\sum_{k=0}^{\infty}F_{-,k}(\hat{\boson}^\dagger)^{k+1} \hat{\boson}^k$,
and $(\hat{\boson}^\dagger)^{k+1} \hat{\boson}^k|\hat{n}=n\rangle
=\frac{n!\sqrt{n+1}}{(n-k)!}|\hat{n}=n+1\rangle$.
The condition 
\[
\begin{aligned}
& \hat{f}_-|\hat{n}=n\rangle
\\
=
&
\left \{\begin{array}{ll}
\sqrt{(2S-n)(n+1)}f(S-n)|\hat{n}=n+1\rangle, & n\leq 2S;\\
0, & n> 2S,
\end{array}\right .
\end{aligned}
\]
becomes

\[
\sum_{k=0}^{n} \frac{n!}{(n-k)!}F_{-,k}
=\left \{\begin{array}{ll}
\sqrt{2S-n}f(S-n), & n\leq 2S;\\
0, & n> 2S.
\end{array}\right .
\]
Define generating function $F_-(t)=\sum_{k=0}^{\infty} F_{-,k}t^k$, 
then $F_-(t)=e^{-t}\sum_{k=0}^{2S}\frac{t^k \sqrt{2S-k}f(S-k)}{k!}$, 
and the projected representation should be 
\[
\begin{aligned}
&
(\hat{f}_-)^\Proj=(\hat{S}_- f(\hat{S}_z))^\Proj= 
\ 
:\hat{\boson}^\dagger F_-(\hat{n}): 
\\
=
&
\ 
:e^{-\hat{n}}\sum_{k=0}^{2S}\frac{\hat{\boson}^\dagger\hat{n}^k}{k!}\sqrt{2S-k}f(S-k):\ .
\end{aligned}
\]
The results for $\hat{S}_-$ and $\hat{S}_-\hat{S}_z+\hat{S}_z\hat{S}_-
=\hat{S}_-(2\hat{S}_z-1)$ 
correspond to $f(S-n)=1$ and $(2(S-n)-1)$, respectively.

Similarly we have the general results for operators changing $S_z$ by $-2$, 
\[
\begin{aligned}
&
(\hat{f}_{--})^\Proj=(\hat{S}_-\hat{S}_- f(\hat{S}_z))^\Proj
\\
=
&
\ 
:e^{-\hat{n}}\sum_{k=0}^{2S}\frac{\hat{\boson}^\dagger\hat{\boson}^\dagger\hat{n}^k}{k!}\sqrt{(2S-k)(2S-k-1)}f(S-k):\ .
\end{aligned}
\]
And this derivation can be extended to the projected HP representation for any single-spin operator. 

\begin{widetext}

\section{Exact bimagnon Hamiltonian for FM states of spin-$1/2$ XXZ model on square lattice}
\label{appsec:BimagnonSpinHamiltonian}

For the spin-$1/2$ XXZ model Eq.~(\ref{equ:H-XXZ-FM}), 
denote the FM ground state (state with $\hat{S}_{i,z}=1/2$) by $|\text{vac}\rangle$. 
Then define the real-space basis for two spin flips, 
$|\Vect{r}_1,\Vect{r}_2\rangle\equiv\hat{S}_{\Vect{r}_1,-}\hat{S}_{\Vect{r}_2,-}|\text{vac}\rangle
=|\Vect{r}_2,\Vect{r}_1\rangle$, note that this state vanishes if $\Vect{r}_1=\Vect{r}_2$ for spin-$1/2$.
The action result of spin Hamiltonian Eq.~(\ref{equ:H-XXZ-FM}) on $|\Vect{r}_1,\Vect{r}_2\rangle$ basis is,
\begin{equation*}
\hat{H}_{\text{FM}}|\Vect{r}_1,\Vect{r}_2\rangle=h(\Vect{r}_1-\Vect{r}_2)|\Vect{r}_1,\Vect{r}_2\rangle
-\frac{J_\perp}{2} \sum_{\Vect{d}} 
(|\Vect{r}_1+\Vect{d},\Vect{r}_2\rangle
+|\Vect{r}_1-\Vect{d},\Vect{r}_2\rangle
+|\Vect{r}_1,\Vect{r}_2+\Vect{d}\rangle
+|\Vect{r}_1,\Vect{r}_2-\Vect{d}\rangle).
\end{equation*}
where 
$\Vect{d}$ runs over the two NN displacement vectors, 
namely $(1,0)$ and $(0,1)$;
the diagonal term $h(\Vect{r}_1-\Vect{r}_2)$ is
\[
h(\Vect{R})
=\left\{\begin{array}{ll}
2h+3J_z, & \Vect{R}\mod (L_x,L_y)\text{ is one of }\Vect{d}\text{ or }-\Vect{d};\\
2h+4J_z, & \text{otherwise and }\Vect{R}\neq (0,0)\mod (L_x,L_y).
\end{array}\right.
\]
For a $L_x\times L_y$ lattice ($N_{\text{site}}=L_x L_y$) with periodic boundary condition, 
lattice momentum can be
$\Vect{k}=(\frac{2\pi}{L_x} n_x, \frac{2\pi}{L_y} n_y)$
with $n_x=0,1,\dots,(L_x-1)$ and $n_y=0,1,\dots,(L_y-1)$. 
Define plane wave state of two spin flips separated by $\Vect{R}$
with center-of-mass momentum $\Vect{k}$ (see \text{e.g.} Ref.~\cite{PhysRevLett.11.336}), 
\begin{equation*}
|\Vect{k};\Vect{R}\rangle
\equiv A_{\Vect{k},\Vect{R}} \sum_{\Vect{r}} |\Vect{r},\Vect{r+R}\rangle \exp[i\Vect{k}\cdot (\Vect{r}+\Vect{R}/2)]=|\Vect{k};-\Vect{R}\rangle,
\end{equation*}
the normalization factor $A_{\Vect{k},\Vect{R}}$ is
\begin{equation*}
A_{\Vect{k},\Vect{R}}= 
\left \{\begin{array}{ll}
(2N_{\text{site}})^{-1/2},& \Vect{R}=(L_x/2,L_y/2),\text{ and }e^{i\Vect{k}\cdot \Vect{R}}=1,\text{ namely }(n_x+n_y)\text{ is even};\\
(2N_{\text{site}})^{-1/2},& \Vect{R}=(L_x/2,0),\text{ and }e^{i\Vect{k}\cdot \Vect{R}}=1,\text{ namely }n_x\text{ is even};\\
(2N_{\text{site}})^{-1/2},& \Vect{R}=(0,L_y/2),\text{ and }e^{i\Vect{k}\cdot \Vect{R}}=1,\text{ namely }n_y\text{ is even};\\
(N_{\text{site}})^{-1/2}, & \text{otherwise}.
\end{array}\right.
\end{equation*}
Note that $|\Vect{k};\Vect{R}\rangle$ vanishes if
$\Vect{R}=(0,0)\mod (L_x,L_y)$, or $\Vect{R}=(L_x/2,L_y/2)\mod (L_x,L_y)$ and $(n_x+n_y)$ is odd,
or $\Vect{R}=(L_x/2,0)\mod (L_x,L_y)$ and $n_x$ is odd,
or $\Vect{R}=(0, L_y/2)\mod (L_x,L_y)$ and $n_y$ is odd.

The action of Hamiltonian on the plane wave states produces
\begin{equation*}
\hat{H}_{\text{FM}}|\Vect{k};\Vect{R}\rangle=
h(\Vect{R})|\Vect{k};\Vect{R}\rangle
-J_\perp
\sum_{\Vect{d}}\cos(\frac{1}{2}\Vect{k}\cdot \Vect{d})
\left (\frac{A_{\Vect{k},\Vect{R}}}{A_{\Vect{k},\Vect{R}+\Vect{d}}}
|\Vect{k};\Vect{R}+\Vect{d}\rangle
+
\frac{A_{\Vect{k},\Vect{R}}}{A_{\Vect{k},\Vect{R}-\Vect{d}}}
|\Vect{k};\Vect{R}-\Vect{d}\rangle
\right ).
\end{equation*}
For given $\Vect{k}$, enumerating $\Vect{R}$ 
over half of the lattice
($\Vect{R}$ and $-\Vect{R}$ correspond to the same plane wave state) 
produces the exact bimagnon Hamiltonian matrix.
Diagonalization of this sparse matrix gets the exact bimagnon energies. 
Note that the ``boundary condition'' for $\Vect{R}$ in $|\Vect{k};\Vect{R}\rangle$ may not be periodic. 
For integer ``winding numbers'' $w_x,w_y$,  
\begin{equation*}
|\Vect{k};\Vect{R}+(w_xL_x,w_yL_y)\rangle=\exp[i (k_xw_xL_x+k_yw_yL_y)/2]|\Vect{k};\Vect{R}\rangle=(-1)^{n_xw_x+n_yw_y}|\Vect{k};\Vect{R}\rangle,
\end{equation*}

If $L_x$ and $L_y$ are both even, 
then for $\Vect{k}=(\pi,\pi)$ 
the above hamitonian is diagonal, 
the exact bimagnon excitation energies are 
$\epsilon_{(\pi,\pi),\Vect{R}}=2h+3J_z$ for $\Vect{R}=(1,0)$ or $(0,1)$,
and
$\epsilon_{(\pi,\pi),\Vect{R}}=2h+4J_z$ for other $\Vect{R}$.
Here the energy $2h+3J_z$ states are bimagnon (quasi-)bound states out of the continuum.

\section{Bimagnon Hamiltonian from bosonic representations of FM XXZ model on square lattice}
\label{appsec:BimagnonBosonHamiltonian}

The two-magnon interaction terms for projected HP representation Eq.~(\ref{equ:FM-Interactions-P})
and the original HP representation Eq.~(\ref{equ:FM-Interactions-HP})
can both be expressed in terms of single magnon operators as follows, 
\begin{equation*}
\hat{H}^{(2)}_{\text{FM}}=\frac{1}{N_{\text{site}}}
\sum_{\Vect{q},\Vect{k},\Vect{k}'}
\frac{1}{4}
V_{\Vect{q},\Vect{k},\Vect{k}'}
\hat{\boson}_{\Vect{k}'}^\dagger
\hat{\boson}_{\Vect{q}-\Vect{k}'}^\dagger
\hat{\boson}_{\Vect{q}-\Vect{k}}^\nd
\hat{\boson}_{\Vect{k}}^\nd.
\end{equation*}
Here $V_{\Vect{q},\Vect{k},\Vect{k}'}$ 
should be symmetrized such that 
$V_{\Vect{q},\Vect{k},\Vect{k}'}=V_{\Vect{q},\Vect{q}-\Vect{k},\Vect{k}'}
=V_{\Vect{q},\Vect{k},\Vect{q}-\Vect{k}'}$.
For original and projected HP representations 
of boson interactions Eq.~(\ref{equ:AFM-Interactions-HP},\ref{equ:AFM-Interactions-P}), 
$V_{\Vect{q},\Vect{k},\Vect{k}'}$ are,
\begin{equation*}
\begin{aligned}
V^\HP_{\Vect{q},\Vect{k},\Vect{k}'}
=
& 
-4D_z-2J_z [\cos(k_x-k'_x)+\cos(q_x-k_x-k'_x)
+(x\to y)]
\\
&
+J_\perp
[\cos(k_x)+\cos(q_x-k_x)
+\cos(k'_x)+\cos(q_x-k'_x)
+(x\to y)],
\\
V^\Proj_{\Vect{q},\Vect{k},\Vect{k}'}
=
& 
-(16SJ_z+4h)\delta_{S,1/2}
-4D_z(1-\delta_{S,1/2})
-2J_z [\cos(k_x-k'_x)+\cos(q_x-k_x-k'_x)
+(x\to y)]
\\
&
+4SJ_\perp
\left (1-\sqrt{1-\frac{1}{2S}}\right )
[\cos(k_x)+\cos(q_x-k_x)
+\cos(k'_x)+\cos(q_x-k'_x)
+(x\to y)].
\end{aligned}
\end{equation*}
Here $(x\to y)$ means replacing the $x$-components of momenta 
by $y$-components in previous terms.

Define basis $|\Vect{q};\Vect{k}\rangle
\equiv A_{\Vect{q},\Vect{k}}^\nd\hat{\boson}_{\Vect{k}}^\dagger\hat{\boson}_{\Vect{q}-\Vect{k}}^\dagger|\text{vac}\rangle
=|\Vect{q};\Vect{q}-\Vect{k}\rangle$,
where 
\[
A_{\Vect{q},\Vect{k}}=\left \{\begin{array}{ll}
1/\sqrt{2}, & \Vect{q}=2\Vect{k}\text{ mod reciprocal vectors};\\
1, & \text{otherwise}.
\end{array}\right.
\]

Then the two-magnon Hamiltonian
$\hat{H}_{\text{LSWT,FM}}+\hat{H}^{(2)}_{\text{FM}}$ acted on this basis produces
\begin{equation*}
(\hat{H}_{\text{LSWT,FM}}+\hat{H}^{(2)}_{\text{FM}})
|\Vect{q};\Vect{k}\rangle
=(\epsilon_{\text{FM},\Vect{k}}+\epsilon_{\text{FM},\Vect{q}-\Vect{k}})|\Vect{q};\Vect{k}\rangle
+\frac{1}{N_{\text{site}}}
\sum_{\Vect{k'}}A_{\Vect{q},\Vect{k}'}A_{\Vect{q},\Vect{k}}V_{\Vect{q},\Vect{k},\Vect{k}'}
|\Vect{q};\Vect{k}'\rangle.
\end{equation*}
For given $\Vect{q}$, enumerating independent $\Vect{k}$
($\Vect{k}$ and $\Vect{q}-\Vect{k}$ correspond to the same basis)
produces the bimagnon Hamiltonian matrix
which can be diagonalized to get the bimagnon energies
for finite size lattices. 
For infinite lattice, additional approximations/truncations are needed to get the bimagnon energies.

\end{widetext}

\pagebreak

\section{Details of NLSWT for honeycomb AFM XXZ model}
\label{appsec:NLSWTHoneycomb}

\begin{figure*}
\includegraphics[scale=0.65]{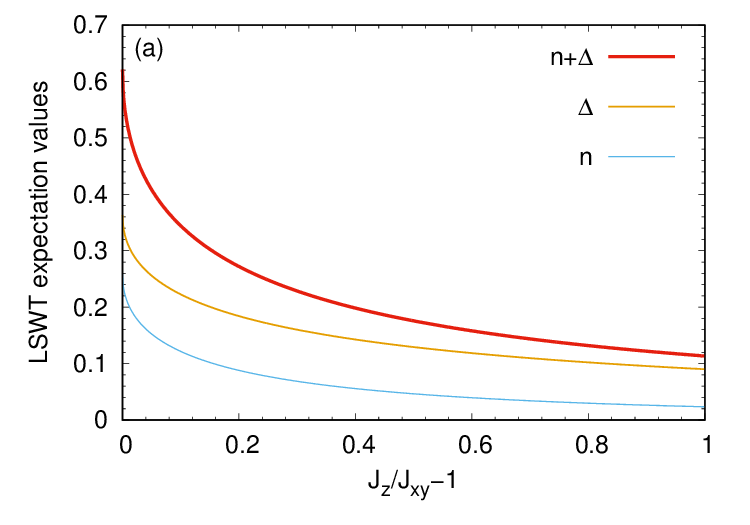} 
\includegraphics[scale=0.65]{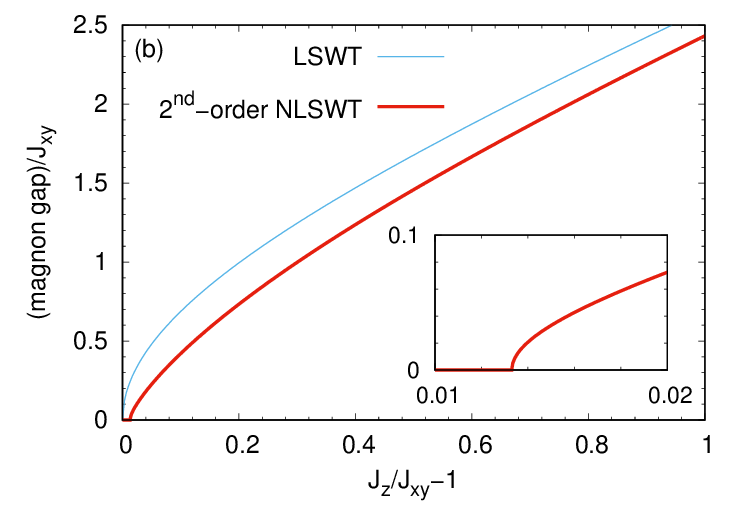}
\caption{
Some LSWT and 2nd-order NLSWT results
for honeycomb AFM easy-axis($J_z\geq J_\perp=J_{xy}=1$) XXZ model Eq.~(\ref{equ:H-AFM}) 
versus the anisotropy $J_z/J_{xy}-1$. 
(a)
LSWT expectation values of boson density $n\equiv \langle \hat{\boson}^\dagger_i\hat{\boson}^\nd_i\rangle$
and NN pairing amplitude $\pairing\equiv \langle \hat{\boson}^\nd_i\hat{\boson}^\nd_j\rangle$
and their sum. 
(b)
The LSWT and 2nd-order NLSWT magnon gap. 
There is a spurious magnon gap closing in NLSWT for $1< J_z/J_{xy} < 1.0133$
(see inset). 
}
\label{fig:NLSWT-Honeycomb}
\end{figure*}

The LSWT and 2nd-order NLSWT calculations for honeycomb AFM XXZ model are essentially the same as Ref.~\cite{PhysRevB.44.11869}.

The LSWT spin wave dispersion is 
\[
\epsilon_{\text{LSWT,AFM},\Vect{k}}
=zS\sqrt{J_z^2-J_{\perp}^2 |\gamma_{\Vect{k}}|^2}
\]
For honeycomb lattice, the coordination number $z=3$, 
and the ``structure factor'' 
$\gamma_{\Vect{k}}=\frac{1}{3}(e^{i k_1}+e^{i k_2}+1)$, 
where 
$k_i\equiv \Vect{k}\cdot \Vect{a}_i$
and the two lattice vectors can be chosen as
$\Vect{a}_1=(a,0)$ and $\Vect{a}_2=(-a/2,a\sqrt{3}/2)$
under Cartesian coordinates
($a$ the the lattice constant). 

The average HP boson density and NN pairing amplitude are
\[
\begin{aligned}
\bar{n} & =\frac{1}{2}\left (\int_{0}^{2\pi}\int_{0}^{2\pi}\frac{\mathrm{d}k_1\mathrm{d}k_2}{4\pi^2}
\frac{zSJ_z}{\epsilon_{\text{LSWT,AFM},\Vect{k}}}-1\right ),
\\
\bar{\pairing} & =\frac{1}{2}\left (\int_{0}^{2\pi}\int_{0}^{2\pi}\frac{\mathrm{d}k_1\mathrm{d}k_2}{4\pi^2}
\frac{zSJ_\perp|\gamma_{\Vect{k}}|^2}{\epsilon_{\text{LSWT,AFM},\Vect{k}}}\right ).
\end{aligned}
\]
Then we can compute the magnon gap in LSWT and 2nd-order NLSWT 
at $\Vect{k}=0$ discussed in main text. 
The results are shown in Fig.~\ref{fig:NLSWT-Honeycomb}. 
Because the large quantum fluctuations in honeycomb models due to its small coordination number, 
the 2nd-order NLSWT magnon gap has a spurious closing near the Heisenberg limit.

\begin{widetext}

\section{Evaluation of expectation values for some projected operators under LSWT ground states of Spin-$1/2$ N\'eel orders}
\label{appsec:LSWTExpectationValue}

The projection operator and projected spin operators Eq.~(\ref{equ:ProjRepS.a}-\ref{equ:ProjRepS.d}) for spin-$1/2$ are explicitly, 
\[
\hat{P}_{1/2}^\nd = \sum_{k=0}^{\infty}\frac{(-1)^k(1-k)}{k!} \hat{\boson}^{\dagger k}\hat{\boson}^{\nd k},
\]
and
\[
\hat{S}_z^\Proj = \frac{1}{2}\sum_{k=0}^{\infty}\frac{(-1)^k(k+1)}{k!} \hat{\boson}^{\dagger k}\hat{\boson}^{\nd k},
\quad
\hat{S}_+^\Proj = \sum_{k=0}^{\infty}\frac{(-1)^k}{k!} \hat{\boson}^{\dagger k}\hat{\boson}^{\nd k+1},
\quad
\hat{S}_-^\Proj = \sum_{k=0}^{\infty}\frac{(-1)^k}{k!} (\hat{\boson}^{\dagger})^{k+1}\hat{\boson}^{\nd k}.
\]

Under LSWT ground state for N\'eel order on bipartite Archimedean lattice, 
the Wick expansion for the expectation values 
of single-spin and two-spin operators on NN bonds contains only 
$\bar{n}\equiv \LSWTexp{\hat{\boson}_i^\dagger\hat{\boson}_i^\nd}$
(which is independent of site $i$),
and 
$\bar{\pairing}\equiv \LSWTexp{\hat{\boson}_i^\nd\hat{\boson}_j^\nd}$
(which is real and uniform on all NN bonds), 
due to the stagger $U(1)$ symmetry.

Then the expectation values for $\hat{P}_{1/2,i}$ and $\hat{S}_{i,z}^\Proj$ are 
\begin{align*}
\LSWTexp{ \hat{P}_{1/2,i} }
&
=\sum_{k=0}^{\infty}(-1)^k(1-k) \bar{n}^{k}
=\frac{1+2\bar{n}}{(1+\bar{n})^2},
\\
\LSWTexp{\hat{S}_{i,z}^\Proj}
&
=\frac{1}{2}\sum_{k=0}^{\infty}(-1)^k(k+1) \bar{n}^{k}
=\frac{1}{2(1+\bar{n})^2}.
\end{align*}

For the LSWT expectation values of
$(\hat{S}_{i,z}^\Proj\hat{S}_{j,z}^\Proj-S^2\hat{P}_{S,i}\hat{P}_{S,j})$, 
$\hat{S}_{i,+}^\Proj\hat{S}_{j,+}^\Proj$, 
and
$\hat{P}_{S,i}\hat{P}_{S,j}$, 
we need the expectation values of $\hat{\boson}_i^{\dagger k_1}
\hat{\boson}_i^{\nd k_1}
\hat{\boson}_j^{\dagger k_2}
\hat{\boson}_j^{\nd k_2}$
and
$\hat{\boson}_i^{\dagger k_1}
\hat{\boson}_i^{\nd k_1+1}
\hat{\boson}_j^{\dagger k_2}
\hat{\boson}_j^{\nd k_2+1}$, which are respectively,  
\begin{equation*}
\begin{aligned}
\LSWTexp{\hat{\boson}_i^{\dagger k_1}
\hat{\boson}_i^{\nd k_1}
\hat{\boson}_j^{\dagger k_2}
\hat{\boson}_j^{\nd k_2}}
=
& 
\sum_{p=0}^{\text{min}(k_1,k_2)}
\bar{\pairing}^{2p}
\bar{n}^{k_1+k_2-2p}
\binom{k_1}{p}\binom{k_1}{p}\binom{k_2}{p}\binom{k_2}{p}
p!p!(k_1-p)!(k_2-p)!
\\
=
&
\sum_{p=0}^{\text{min}(k_1,k_2)}
\bar{\pairing}^{2p}
\bar{n}^{k_1+k_2-2p}
\frac{k_1!k_1!k_2!k_2!}{p!p!(k_1-p)!(k_2-p)!},
\end{aligned}
\end{equation*}
and
\begin{equation*}
\begin{aligned}
\LSWTexp{\hat{\boson}_i^{\dagger k_1}
\hat{\boson}_i^{\nd k_1+1}
\hat{\boson}_j^{\dagger k_2}
\hat{\boson}_j^{\nd k_2+1}}
=
& 
\sum_{p=0}^{\text{min}(k_1,k_2)}
\bar{\pairing}^{2p+1}
\bar{n}^{k_1+k_2-2p}
\binom{k_1}{p}\binom{k_1+1}{p+1}\binom{k_2}{p}\binom{k_2+1}{p+1}
p!(p+1)!(k_1-p)!(k_2-p)!
\\
=
&
\sum_{p=0}^{\text{min}(k_1,k_2)}
\bar{\pairing}^{2p+1}
\bar{n}^{k_1+k_2-2p}
\frac{k_1!(k_1+1)!k_2!(k_2+1)!}{p!(p+1)!(k_1-p)!(k_2-p)!}.
\end{aligned}
\end{equation*}

Then the expectation value of 
the $J_z$ terms $(\hat{S}_{i,z}^\Proj\hat{S}_{j,z}^\Proj-S^2\hat{P}_{S,i}\hat{P}_{S,j})$ is
\begin{equation*}
\begin{aligned}
& \LSWTexp{ \hat{S}_{i,z}^\Proj\hat{S}_{j,z}^\Proj-S^2\hat{P}_{S,i}\hat{P}_{S,j} }
\\
=
&
\frac{1}{4}\sum_{k_1=0}^{\infty}
\sum_{k_2=0}^{\infty}\frac{ (-1)^{k_1}(-1)^{k_2}[(k_1+1)(k_2+1)-(1-k_1)(1-k_2) ] } {k_1! k_2!}
\LSWTexp{\hat{\boson}_i^{\dagger k_1}
\hat{\boson}_i^{\nd k_1}
\hat{\boson}_j^{\dagger k_2}
\hat{\boson}_j^{\nd k_2}}
\\
=
&
\frac{1}{2}\sum_{k_1=0}^{\infty}
\sum_{k_2=0}^{\infty}
\sum_{p=0}^{\text{min}(k_1,k_2)}
\bar{\pairing}^{2p}
\bar{n}^{k_1+k_2-2p}
\frac{ (-1)^{k_1} (-1)^{k_2} k_1! k_2! (k_1+k_2) }{p!p!(k_1-p)!(k_2-p)!}
\\
=
&
\sum_{p=0}^{\infty}
\frac{
\bar{\pairing}^{2p}
\bar{n}^{-2p}
}{p!p!}
\left (\sum_{k_1=p}^{\infty}
\bar{n}^{k_1}
\frac{ (-1)^{k_1}[(k_1+1)!-k_1!] }{(k_1-p)!}
\right )
\left (\sum_{k_2=p}^{\infty}
\bar{n}^{k_2}
\frac{ (-1)^{k_2}k_2! }{(k_2-p)!}
\right )
\\
= 
&
\sum_{p=0}^{\infty}
\frac{
\bar{\pairing}^{2p}
\bar{n}^{-2p}
}{p!p!}
\cdot 
(-1)^p \bar{n}^p \left (\frac{(p+1)!}{(1+\bar{n})^{p+2}}-\frac{p!}{(1+\bar{n})^{p+1}} \right )
\cdot
\frac{(-1)^p \bar{n}^p p!}{(1+\bar{n})^{p+1}}
\\
=
&
\frac{1}{4}\sum_{p=0}^{\infty}
\bar{\pairing}^{2p}
(1+\bar{n})^{-2p-3} 
(p-\bar{n})
=-\frac{\bar{n}+\bar{n}^2-\bar{\pairing}^2}{[(1+\bar{n})^2-\bar{\pairing}^2]^2}
\approx 
-\bar{n}+3\bar{n}^2+\bar{\pairing}^2+\dots\ .
\end{aligned}
\end{equation*}

Similarly $\LSWTexp{ \hat{P}_{S,i}\hat{P}_{S,j} }$ is
\begin{equation*}
\begin{aligned}
& \LSWTexp{ \hat{P}_{S,i}\hat{P}_{S,j} }
\\
=
& 
\sum_{k_1=0}^{\infty}
\sum_{k_2=0}^{\infty}\frac{ (-1)^{k_1}(-1)^{k_2}(1-k_1)(1-k_2) } {k_1! k_2!}
\LSWTexp{\hat{\boson}_i^{\dagger k_1}
\hat{\boson}_i^{\nd k_1}
\hat{\boson}_j^{\dagger k_2}
\hat{\boson}_j^{\nd k_2}}
\\
=
&
\sum_{p=0}^{\infty}
\bar{\pairing}^{2p} \cdot 
\left (\frac{(p+1)}{(1+\bar{n})^{p+2}}-\frac{2}{(1+\bar{n})^{p+1}} \right )^2
\\
=
&
\frac{(1+\bar{n})^2+\bar{\pairing}^2}{[(1+\bar{n})^2-\bar{\pairing}^2]^3}
-\frac{4(1+\bar{n})}{[(1+\bar{n})^2-\bar{\pairing}^2]^2}
+\frac{4}{(1+\bar{n})^2-\bar{\pairing}^2}
\approx 1-2\bar{n}^2+\dots\ .
\end{aligned}
\end{equation*}

And the expectation value of $J_\perp$ term 
$ \hat{S}_{i,x}^\Proj\hat{S}_{j,x}^\Proj
-\hat{S}_{i,y}^\Proj\hat{S}_{j,y}^\Proj
=\frac{1}{2}
(\hat{S}_{i,-}^\Proj\hat{S}_{j,-}^\Proj
+\hat{S}_{i,+}^\Proj\hat{S}_{j,+}^\Proj )
$ 
equals to
\begin{equation*}
\begin{aligned}
\LSWTexp{ \hat{S}_{i,+}^\Proj\hat{S}_{j,+}^\Proj }
=
&
\sum_{k_1=0}^{\infty}
\sum_{k_2=0}^{\infty}\frac{ (-1)^{k_1}(-1)^{k_2} } {k_1! k_2!}
\LSWTexp{\hat{\boson}_i^{\dagger k_1}
\hat{\boson}_i^{\nd k_1+1}
\hat{\boson}_j^{\dagger k_2}
\hat{\boson}_j^{\nd k_2+1}}
\\
=
&
\sum_{k_1=0}^{\infty}
\sum_{k_2=0}^{\infty}
\sum_{p=0}^{\text{min}(k_1,k_2)}
\bar{\pairing}^{2p+1}
\bar{n}^{k_1+k_2-2p}
\frac{ (-1)^{k_1}(k_1+1)! (-1)^{k_2}(k_2+1)! }{p!(p+1)!(k_1-p)!(k_2-p)!}
\\
=
&
\sum_{p=0}^{\infty}
\frac{
\bar{\pairing}^{2p+1}
\bar{n}^{-2p}
}{p!(p+1)!}
\left (\sum_{k_1=p}^{\infty}
\bar{n}^{k_1}
\frac{ (-1)^{k_1}(k_1+1)! }{(k_1-p)!}
\right )
\left (\sum_{k_2=p}^{\infty}
\bar{n}^{k_2}
\frac{ (-1)^{k_2}(k_2+1)! }{(k_2-p)!}
\right )
\\
= 
&
\sum_{p=0}^{\infty}
\frac{
\bar{\pairing}^{2p+1}
\bar{n}^{-2p}
}{p!(p+1)!}
\cdot [(-1)^p \bar{n}^p (1+\bar{n})^{-p-2} (p+1)!]^2
\\
=
&
\sum_{p=0}^{\infty}
\bar{\pairing}^{2p+1}
(1+\bar{n})^{-2p-4} 
(p+1)
=\frac{\bar{\pairing}}{[(1+\bar{n})^2-\bar{\pairing}^2]^2}
\approx \bar{\pairing}-4\bar{n}\bar{\pairing}+\dots\ .
\end{aligned}
\end{equation*}
Note that the convergence of these infinite series requires that 
$|\bar{n}|<1$ and $|\bar{\pairing}|< 1$. 

\end{widetext}

\bibliography{ProjectedHPBoson}

\end{document}